\documentclass[amsmath,amssymb,floatfix,prl,epsf,twocolumn,superscriptaddress]{revtex4-1}

\usepackage{amsmath,amssymb,bm,graphicx,url,epsfig}
\usepackage[ansinew]{inputenc}
\usepackage{bm}
\usepackage{color}

\newcommand{\be}{\begin{equation}}
\newcommand{\ee}{\end{equation}}
\newcommand{\bes}{\begin{equation}\begin{split}}
\newcommand{\ees}{\end{split}\end{equation}}
\newcommand{\bea}{\begin{eqnarray}}
\newcommand{\eea}{\end{eqnarray}}
\newcommand{\nn}{\nonumber}

\def\beq{\begin{equation}}
\def\eeq{\end{equation}}
\def\bea{\begin{eqnarray}}
\def\eea{\end{eqnarray}}
\begin{document}

\title{Absence of Bose Condensation on Lattices with Moat Bands}



\author{Tigran A. Sedrakyan}

\affiliation{William I. Fine Theoretical Physics Institute and Department of Physics,  University of Minnesota, Minneapolis, Minnesota 55455, USA}

\author{Leonid I. Glazman}

\affiliation{Department of Physics, Yale University, New Haven, Connecticut 06520, USA}

\author{Alex Kamenev}

\affiliation{William I. Fine Theoretical Physics Institute and Department of Physics,  University of Minnesota, Minneapolis, Minnesota 55455, USA}

\begin{abstract}
We study hard-core bosons on a class of frustrated lattices with the lowest Bloch band having a
degenerate minimum along a closed contour, the moat, in the reciprocal space. We show that at small density the ground
state of the system is given by a non-condensed state, which may be viewed as a state of fermions
subject to Chern-Simons gauge field. At fixed density of bosons, such a state exhibits domains of
incompressible liquids. Their fixed densities are given by fractions of the reciprocal-space area enclosed
by the moat.

\end{abstract}

\maketitle



Though initially introduced for an ideal Bose gas, notion of Bose-Einstein condensation\cite{Einstein} (BEC) goes far beyond the non-interacting case and describes, e.g., superfluidity in such strongly correlated liquid as $^4$He\cite{Pines}.
Condensation remains advantageous even at strong interaction, as condensed particles avoid exchange interaction, thus reducing the average potential energy. Within this picture, elementary excitations, the quasiparticles\cite{Landau}, exhibit Bose statistics and gapless sound-like spectrum (for neutral superfluids). These predictions found countless confirmations in a diverse range of systems from $^4$He liquid to cold gases of alkali atoms\cite{Leggett,Pitaevskii}.

\begin{figure}
\centerline{\includegraphics[width=85mm,angle=0,clip]{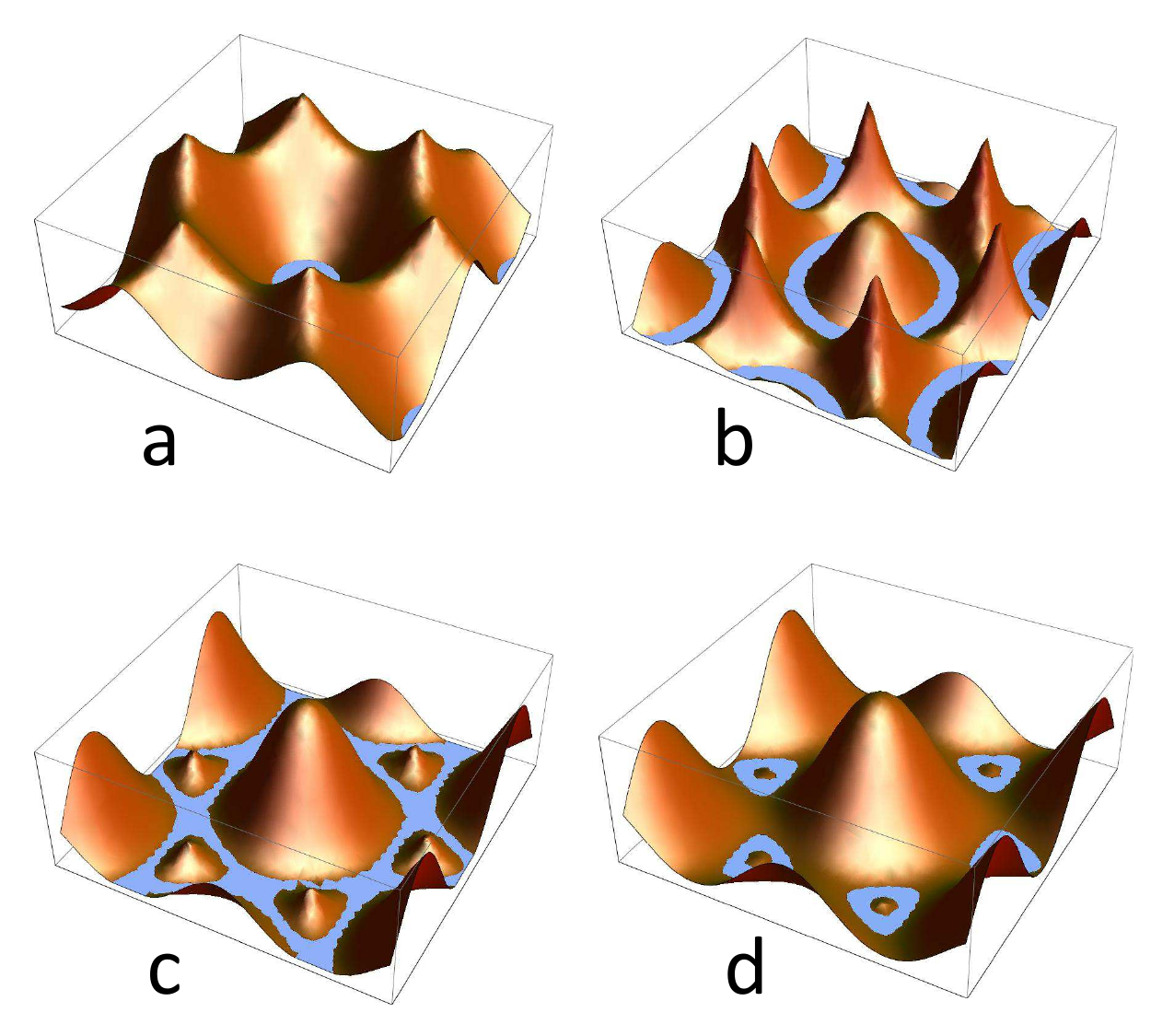}}
\caption{(Color online) Lowest energy band of the honeycomb lattice with $t_2/t_1 =0$ (a), $0.3$ (b), $0.5$ (c), $0.9$ (d). The minimal
energy contour -- the moat ${\cal M}$, is shown in light gray (blue). }
\label{fig-bands}
\end{figure}

The fundamental question is whether BEC ground state with bosonic quasiparticle excitations is the universal faith of any non-crystalline Bose substance.  The goal of this paper is to present  an alternative to this paradigm.
To this end we discuss bosonic liquids in a family of 2D lattices, whose band structure exhibits an energy minimum along a closed line -- the {\em moat}, in the Brillouin zone, Fig.~\ref{fig-bands}. The simplest example of the moat lattice is given by graphene's honeycomb lattice with nearest and next-nearest hopping\cite{exp1,exp2,exp3,exp4,varney3}. With no interactions the ground state is highly degenerate as bosons may condense in any state along the moat as well as in any linear superposition of such states. One may expect that interactions remove the degeneracy and select
a unique ground state.  

In this paper we show that at small filling factors the ground state does not exhibit BEC. Instead of selecting a single macroscopically occupied state, it involves {\em all} states along the moat and its vicinity, each one being only singly occupied. Such a state does not break the underlying $U(1)$ symmetry, although does break the time-reversal invariance.    Moreover, the elementary excitations are not bosons, but rather fermions. At small enough filling fractions, their spectrum is gapped.  As a result, moat lattices provide an example of dramatic departure from the ``BEC + Landau quasiparticles'' paradigm for Bose liquids.

A useful insight in the physics of the moat lattices comes from analogy with the fractional quantum Hall effect (FQHE). There too, the macroscopic degeneracy of the ground state is lifted by the interactions. It results in incompressible (i.e. gapped) states,  when the electron density is an odd integer fraction of the lowest Landau level maximum occupation. There is a similar phenomenology associated with the lifting of degeneracy  in the moat lattices. The characteristic particle density is given by the area  ${\cal A_M}$ of the reciprocal space, enclosed by the moat. For a {\em fractional}  filling of the form
\be
\nu_l = \frac{\cal A_M}{2 l +1+\kappa}\,, \quad\quad\quad  l=0,1,\ldots\,,
                                                                              \label{eq-density-quantized}
\ee
the bosonic ground state is incompressible, here the reciprocal area is normalized to that of the Brillouin zone. 
Index $\kappa$ is related to the reciprocal space Berry phase and is given by $\kappa=0$ if the moat encircles $\Gamma$ point, Fig.~\ref{fig-bands}b, and $\kappa=1$ for moat encircling $K$ and $K'$ points, Fig.~\ref{fig-bands}d.   
For a generic lattice filling $\nu < 1/2$, such that $\nu_{l-1}<\nu<\nu_l$, the system breaks into incompressible domains
with fillings $\nu_{l-1}$ and $\nu_l$.

\begin{figure}
\centerline{\includegraphics[width=85mm,angle=0,clip]{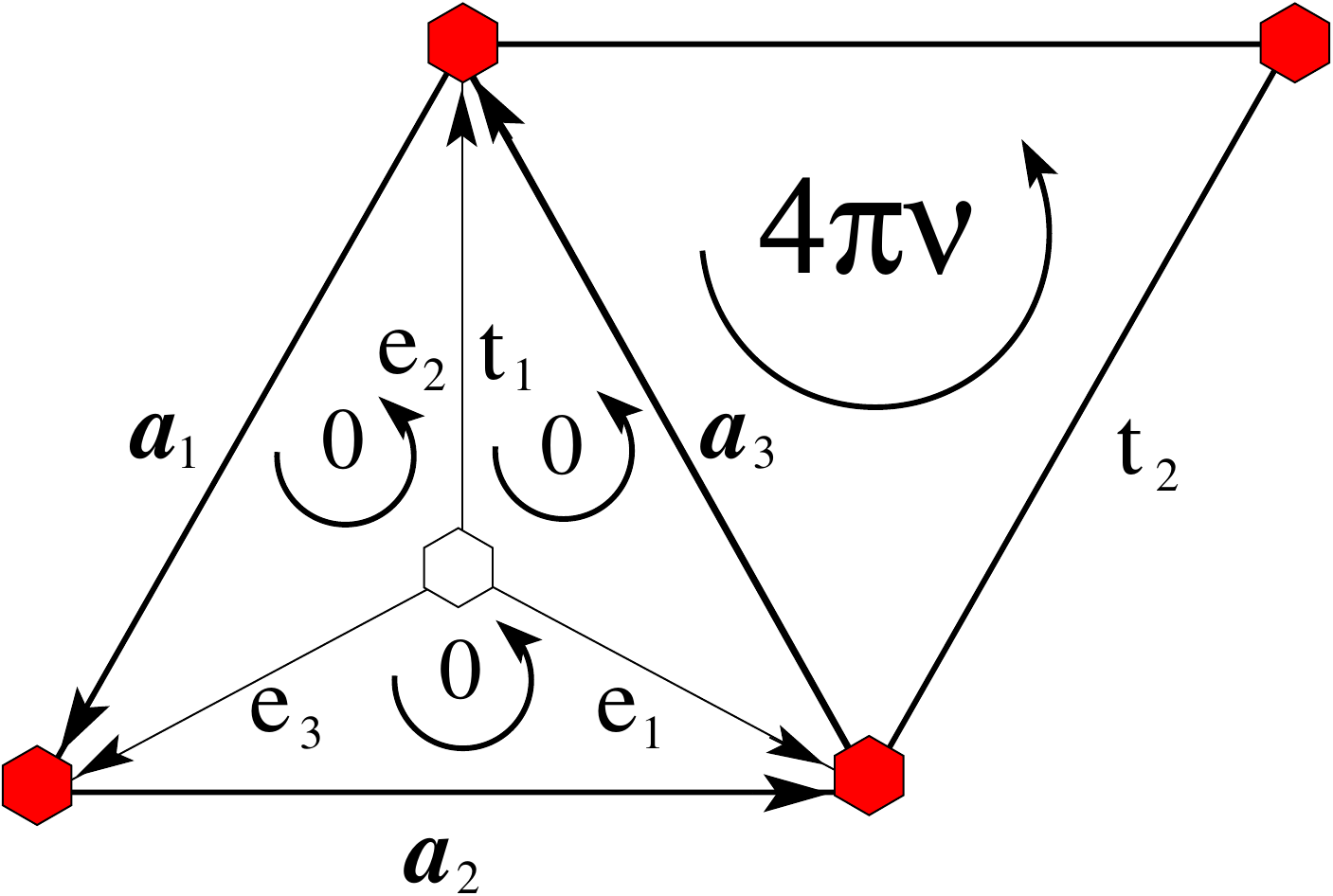}}
\caption{(Color online) The unit cell of honeycomb lattice with lattice vectors ${\bm a}_i$ and ${\bf e}_i$, $i=1,2,3$. Full (empty) cites
belong to the sublattice $A$ ($B$). Total Chern-Simons flux is a combination of (i) $\pi \nu$ fluxes through each of the triangles; (ii) phases $\exp{(-i\pi\nu)}$ attached to sides of the full regular triangle and   $\exp{(i\pi\nu)}$ attached to the sides of the empty regular triangle. This arrangement of phases corresponds to the Haldane modulation of phases with staggered $\phi_H= -\Phi/6= -2\pi \nu/3$ (see main text).
}
\label{fig-vectors}
\end{figure}

Mathematically the moat bands appear, when the lattice Hamiltonian acquires a polynomial structure of the form      
\bea
\label{H1}
\hat H = t_1 \hat T  + t_2 \hat T^2\, , \quad\quad \hat T =
\Bigg(
\begin{array}{cc}
 0& \hat G\\
 \hat G^\dagger&0
\end{array}
\Bigg)\,,
\eea
where the matrix structure is in A/B sublattice space and  $t_1$ and $t_2$ are nearest and next-nearest  hopping, correspondingly. For the case of honeycomb lattice, Fig.~\ref{fig-vectors}, $\hat G= G_{\bf k} = \sum_{j=1,2,3}  e^{i{\bf k}\cdot {\bf e}_j}$ with the three lattice vectors ${\bf e}_j$ connecting a site of sublattice A with three nearest neighbors of sublattice B. The Hamiltonians of the form (\ref{H1}) are not limited, though, to the honeycomb lattice. A generic oblique
lattice with three distinct nearest and three distinct next-nearest hopping integrals is described by Eq.~(\ref{H1}), if {\em two}
conditions are imposed on six  hopping constants \cite{foot1} (variety of other lattices give rise to Hamiltonians of the form (\ref{H1})).

The two energy bands of the Hamiltonian (\ref{H1}) are given by $E_{\bf k}^{(\mp)} =\mp |t_1| |G_{\bf k}| + t_2 |G_{\bf k}|^2 $.
The lowest energy band $E_{\bf k}^{(-)}$ exhibits a degenerate minimum along the contour ${\cal M}$ -- the moat, in the reciprocal space given by $|G_{\bf k}|=|t_1|/2 t_2$. For the honeycomb lattice this condition\cite{foot2} is satisfied for $t_2>|t_1|/6$,  Fig.~\ref{fig-bands}. A similar dispersion relation appears in the context of particles with isotropic  Rashba spin-orbit coupling \cite{we,Rashba,mishch,ian,brandon,gs,berg}.

The issue of Bose condensation for particles with such a dispersion relation is a non-trivial one. 
On the non-interacting level there is no transition at any finite temperature. This is due to the square root, $(E-E_{\cal M})^{-1/2}$, divergence of the single particle DOS near the bottom of the band. Such behavior of DOS highlights similarities with one-dimensional systems, where the ground state of strongly repulsive bosons is given  by the Tonks-Girardeau gas of free fermions
\cite{tonks,girardeau,LiebLiniger,yang,gaudin}. Here we show that the effective fermion picture describes the ground state
of hardcore bosons on  2D moat lattices as well.  
An important observation\cite{we} is that the chemical potential of fermions with the dispersion relation of  Fig.~\ref{fig-bands} scales as $\mu_F\propto \nu^2$ at small enough filling factors $\nu\ll 1$ (this is a consequence of the divergent DOS). 
On the other hand, for BEC  in one of the states along the moat ${\cal M}$, the chemical potential scales as $\mu_B\propto \nu$, due to on-site repulsion  (notice that the latter does not affect the fermionic energy, because of the Pauli exclusion).  
One thus concludes that a small enough filling $\nu$ the fermionic state is energetically favorable over BEC.  

\begin{figure}
\centerline{\includegraphics[width=85mm,angle=0,clip]{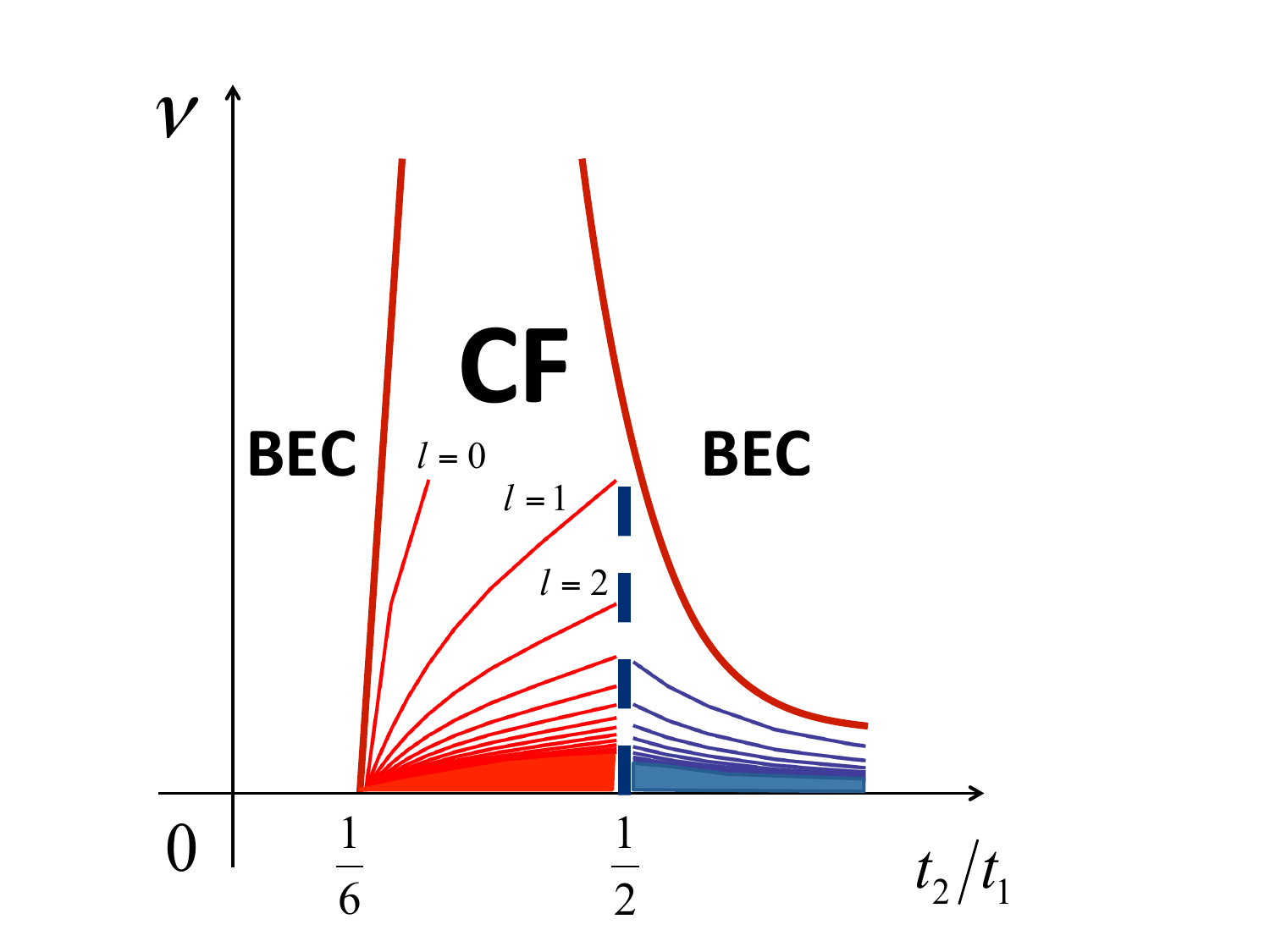}}
\caption{(Color online) Phase diagram of hard-core bosons on a honeycomb lattice. CF and BEC are  composite fermion
and Bose condensate states respectively. Also shown  incompressible states with fractionally quantized filling fractions $\nu_l$.  }
\label{fig-phase}
\end{figure}

To build a fermionic state of Bose particles one may use Chern-Simons flux attachments familiar in the context of FQHE\cite{Jain,lopez,Halperin,shan,read}. This leads to composite fermions (CF) subject to a dynamic magnetic field produced by the attached flux tubes. Following FQHE ideas, one may treat the latter in the mean-field approximation by substituting on-site density operators by their expectation values. In the context of FQHE this leads to a uniform magnetic filed, which partially  compensates for the external one. The lattice version of this procedure is somewhat more subtle, however. Since the particles (and thus the fluxes, attached to them) are confined to stay on the lattice sites, 
a uniform lattice filling $\nu$ does {\em not} translate into a uniform magnetic field. As we explain below, it rather leads to a uniform magnetic flux $4\pi \nu$ per unit cell superimposed with a {\em staggered} Haldane\cite{Haldane} flux arrangement.   At small filling factors, $\nu \ll 1$, the corresponding Hofstadter spectrum consists of quantized Landau levels, separated 
by  cyclotron gaps. The latter protects the ground state from divergent fluctuation correction, rendering (local) stability of the mean-field ansatz. The corresponding phase diagram is schematically depicted in Fig.~\ref{fig-phase}.

To quantify these ideas we start from the Hamiltonian, written in terms of bosonic creation and annihilation operators $b_{\bf r}^{\dagger}$, $b_{\bf r}$, 
which commute at different cites,  $[b^{\pm}_{\bf r},b_{\bf r'}]=0,\; {\bf r}\neq {\bf r'} $, and fulfill the 
 hard-core condition
$\left(b_{\bf r}^{\dagger}\right)^2=\left(b_{\bf r}\right)^2=0$.
For, {\sl e.g.}, honeycomb lattice the Hamiltonian takes the form:
\be
\label{H-XY}
H= t_1 \sum_{{\bf r},j}b^\dagger_{\bf r}b_{{\bf r}+{\bf e}_j}+
t_2 \sum_{{\bf r},j}b^\dagger_{\bf r}b_{{\bf r}+{\bf a}_j}+   H.c. 
\ee
where the vectors ${\bf e}_j$ and ${\bf a}_j,\; j=1,2,3$ are shown in Fig.~\ref{fig-vectors}.
Chemical potential, $\mu$, is related to the average on-site occupation $\nu$
through an equation of state.

Motivated by the observation that the fermionic chemical potential is lower than that of BEC, we proceed with the Chern-Simons transformation\cite{Jain, lopez, Halperin,shan,read}. To this end we write the bosonic
operators as
\bea
\label{CS}
b_{\bf r}^{(\dagger)}= c_{\bf r}^{(\dagger)}\, e^{\pm i \sum_{{\bf r'}\neq {\bf r}} \arg[{\bf r}-{\bf \tilde r}] n_{\bf \tilde r}}\,,
\eea
where the summation runs over all sites of the lattice. Since the bosonic operators on different sites commute, the newly defined operators $c_{\bf r}$ and $c_{\bf r}^{\dagger}$  obey fermionic commutation relations. Also notice that the number operator is given by $n_{\bf r} = c_{\bf r}^\dagger c_{\bf r}$.  Upon transformation
(\ref{CS})  hopping terms of the Hamiltonian (\ref{H-XY}) acquire phase factors  $ e^{ i \sum_{{\bf \tilde r}} \phi_{{\bf \tilde r}, {\bf r}, {\bf r'} }n_{{\bf \tilde r}} }$, where $\phi_{{\bf \tilde r}, {\bf r}, {\bf r'} }$
is an angle at which the link $\langle{\bf r}, {\bf r'}\rangle$ is seen from the lattice site ${\bf \tilde r}$. In terms of the fermionic operators the Hamiltonian (\ref{H-XY})  reads as
\bea
\label{H2}
H&=&  t_1 \sum_{{\bf r},j}  c^\dagger_{\bf r} c_{{\bf r}+{\bf e}_j} e^{i \sum_{\bf \tilde r} 
 \phi_{{\bf \tilde r}, {\bf r}, {\bf r}+\bf {e}_j } n_{\bf \tilde r} } \nn \\
&+&
t_2 \sum_{{\bf r},j} c^\dagger_{\bf r} c_{{\bf r}+{\bf a}_j} e^{i \sum_{{\bf \tilde r}} \phi_{{\bf \tilde r}, {\bf r}, {\bf r}+{\bf a}_j }n_{\bf \tilde r}} +H.c. 
\eea
Notice that the hard-core condition is taken care of by the  Pauli principle and thus fermions may be considered as non-interacting.  Using the expression for $ \phi_{{\bf \tilde r}, {\bf r}, {\bf r}' }$, one can directly check that $  \sum_{{\bf \tilde r}} \phi_{{\bf \tilde r}, {\bf r}, {\bf r}+{\bf e}_i }n_{{\bf \tilde r}}
- \sum_{{\bf \tilde r}} \phi_{{\bf \tilde r}, {\bf r}, {\bf r}+{e}_j }n_{{\bf \tilde r}}= \sum_{{\bf \tilde r}} \phi_{{\bf \tilde r}, {\bf r}, {\bf r}+{\bf e}_{i}-{\bf e}_{j} }n_{{\bf \tilde r}}$, for any two vectors ${\bf e}_{i/j}$, $i,j=1,2,3$ shown in Fig.~\ref{fig-vectors}. The right hand side of this equation can be identified with
the phase acquired by the next-nearest-neighbor hopping term along ${\bf a}_l={\bf e}_{i}-{\bf e}_{j}$, while the left hand side represents the phase of two consecutive nearest-neighbor (NN) hops along vectors ${\bf e}_{i}$ and $-{\bf e}_{j}$.  
As a result, the Hamiltonian (\ref{H2}) retains the algebraic structure of Eq.~(\ref{H1}), where operator $\hat T$ describes fermions in NN graphene lattice subject to CS fluxes.  

To analyze the consequences of these phase factors we adopt the mean-field ansatz\cite{Jain, Halperin}, $n_{\bf \tilde r}\approx \langle n_{\bf \tilde r} \rangle\equiv \nu$. This substitutes fluctuating CS phases with an external magnetic filed, carrying flux $\Phi=4\pi \nu$ per unit cell (two cites, each with the occupation $\nu$ and $2\pi$ flux per particle). While NN hoping operator $\hat T$ is sensitive only to this total flux, the next-NN operator $\hat T^2$  implies that 
the magnetic filed exhibits Haldane modulation\cite{Haldane} within the unit cell.  Indeed  the phase factor,  corresponding to a link ${\langle{\bf r r' }\rangle}$ is  
$\varphi_{\bf r r' }= \sum_{{\bf \tilde r}\neq {\bf r, r' }} \phi_{\bf \tilde r, r r' } \nu +( \arg[{\bf r}-{\bf  r'}]-\arg[{\bf r'}-{\bf  r}])\nu $.
For a counterclockwise travel along any elemental (i.e. not encircling any lattice points) triangle, the first term here accumulates the net phase $\pi\nu$. The second term brings phase $-\pi\nu$ for small $120^\circ$ triangles and phase
$3\pi\nu$ for large equilateral triangle,  Fig.~\ref{fig-vectors}. As a result, the entire flux $\Phi$ is concentrated into a half 
of the unit cell -- the large empty triangle.  This corresponds to Haldane modulation\cite{Haldane} with the staggering 
parameter $\phi_H=-\Phi/6$,  {\em superimposed} with the uniform flux $\Phi$. Notice, that only such configuration of fluxes 
results in the algebraic Hamiltonian  (\ref{H1}),  while, e.g., a constant magnetic field does {\em not} admit representation (\ref{H1}).


\begin{figure}
\centerline{\includegraphics[width=85mm,angle=0,clip]{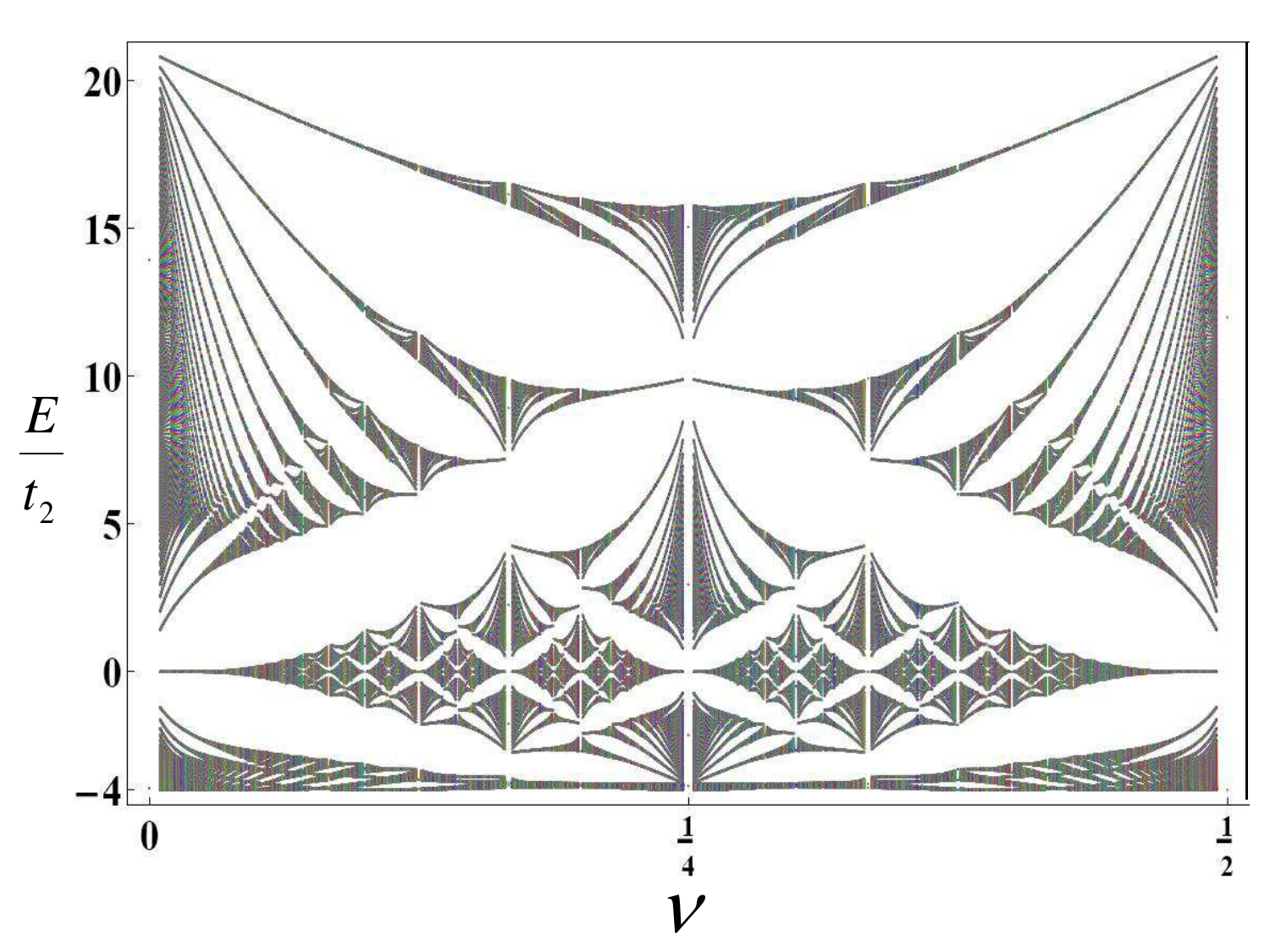}}
\caption{Hofstadter energy spectrum vs. filling fraction $\nu\in [0,1/2]$, for $t_2=t_1/4$, {\sl i.e.} moat $\mathcal{M}$ is
around the $\Gamma$ point.  Notice that the bottom of the Hofstadter spectrum is flat, which is a consequence of
the fact that all Landau levels exhibit minima at the same energy  $E=-t_1^2/4t_2$.}
\label{fig-Hofstadter}
\end{figure}

\begin{figure}
\centerline{\includegraphics[width=85mm,angle=0,clip]{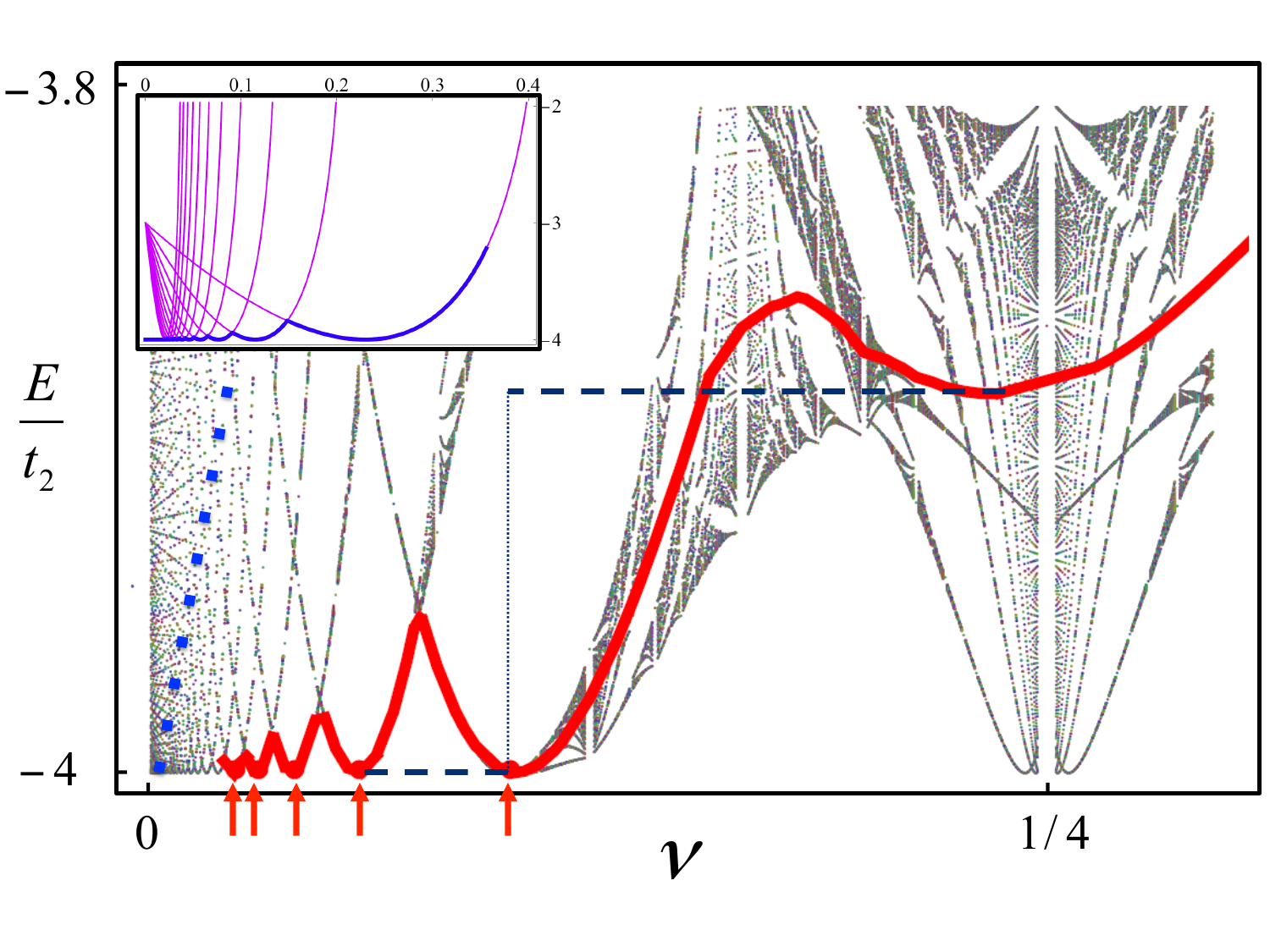}}
\caption{(Color online) Bottom part of Fig.~\ref{fig-Hofstadter}. Thick (red) line represents the ground state energy per particle, $E_{GS}(\nu)$,  obtained numerically from the Hofstadter energy spectrum. Arrows  show fractionally quantized filling fractions (\ref{eq-density-quantized}). Dashed line is the macroscopic chemical potential exhibiting jumps at the fractionally quantized filling fractions. Dotted line is the chemical potential of the Bose condensed state. Inset: Semiclassical Landau levels as functions of the filling fraction.}
\label{fig-Hofstadter1}
\end{figure}

This algebraic structure (\ref{H1}) greatly simplifies spectral problem by reducing it to diagonalization of the NN operator $\hat T$. As mentioned above, the latter is sensitive only to the total flux $\Phi$, but {\em not}  to the
staggered component $\phi_H$.  At small filling factors (i.e. magnetic fields) its spectrum may be analyzed in the semiclassical  approximation\cite{Onsager}.
Accordingly, the eigenvalues of $\hat T$, denoted as 
$G_l(\Phi)$, where $l=0,1,\ldots$, can be found by: (i) considering the
constant energy contours $|G_{\bf k}|=\mathrm{const}=G$ of the bare operator in the reciprocal ${\bf k}$-space, and (ii) identifying  $G_l(\Phi)$ with energy $G$ of contours  having  normalized reciprocal area  
\bea
\label{Onsager}
{\cal A}_l=  \left(l+{1\over 2}-\kappa \right) \frac{\Phi}{2\pi}\,,
\eea
where  $2\pi\kappa$ is  the Berry phase\cite{berry,fuchs}. Finally,
the spectrum of the Hamiltonian (\ref{H1}), which describes the lattice subject to the uniform magnetic flux $\Phi$ {\em and} Haldane modulation $\phi_H=-\Phi/6$, is found in terms of $G_l(\Phi)$ as
\bea
\label{LL}
E_l(\Phi)=-t_1 G_l(\Phi) + t_2 \big[G_l(\Phi)\big]^2  \, .
\eea
Since we have attached exactly one flux quantum per fermion, all states at the lowest Landau level (LLL) are occupied. As a result, the many-body ground state energy follows  LLL. The peculiarity of the moat dispersion is that LLL is not necessarily $l=0$ one, but a level with $l\approx  {\cal A_M}/\nu$, 
see inset in  Fig.~\ref{fig-Hofstadter1}. Indeed, Landau levels (\ref{LL}) are non-monotonic functions of flux. They reach the minimum at $G=t_1/2t_2$, i.e. exactly at the  very bottom of the moat. Recalling that $\Phi=4\pi\nu$, one obtains the set of the filling factors $\nu_l$, Eq.~(\ref{eq-density-quantized}), where LLL (and thus the ground state energy)   reaches its minima. As an illustration, consider  the moat closely encircling $K$ and $K'$ points, Fig.~\ref{fig-bands}d. In this case  $G_{\bf k}\approx 3|k|/2$ and $\kappa=1/2$\cite{Haldane}, from Eq.~(\ref{Onsager}) one finds $G_l(\Phi)=\sqrt{ \sqrt{3}\Phi l}$, leading to $E_l(\Phi)=- t_1\sqrt{ \sqrt{3}\Phi l} + t_2\sqrt{3}\Phi l$. The non-monotonic dependence on $\Phi$ is evident.

To go beyond the semiclassical approximation we consider the Hofstadter problem on the lattice, including Haldane modulation.
For a rational flux  $\Phi=4\pi p/q$ ($p$ and $q$ are positive integers) diagonalization of the operator $\hat T$
reduces to Harper equation,  which can be analyzed numerically. For such fluxes the spectrum splits onto
$q$ non-overlapping subbands, labeled by $m=1,2,\ldots q$.
The corresponding spectrum $E_{m,{\bf k}}(\Phi)$, Fig.~\ref{fig-Hofstadter}, acquires the form of  the Hofstadter butterfly\cite{butterfly}.
Notice the flatness of the lower edge of the spectrum, which reflects the divergent DOS at this energy.
Fig.~\ref{fig-Hofstadter1} amplifies the lowest part of the Hofstadter spectrum. Non-monotonic Landau levels, Eq.~(\ref{LL}), are clearly visible at small filling fractions.  The ground state energy per particle 
$E_{GS}(\nu)= \frac{q}{N p} \sum_{m=1}^p \sum_{\bf k}^{N/q} E_{m,{\bf k}}(4\pi p/q)$, 
where $N$ is number of lattice sites, is shown  In Fig.~\ref{fig-Hofstadter1}.  For small filling fractions it closely follows the semiclassical LLL (\ref{LL}), exhibiting the minima at the fractionally quantized filling fractions  $\nu_l$, Eq.~(\ref{eq-density-quantized}).
Due to Maxwell phase separation rule, this leads to the  macroscopic chemical potential of the staircase shape with the jumps at the fractionally quantized filling fractions  (\ref{eq-density-quantized}), see Fig.~\ref{fig-Hofstadter1}. The flat regions of the staircase imply phase separation into domains with fillings $\nu_l$ and $\nu_{l+1}$.

As seen in Fig.~\ref{fig-Hofstadter1},  for $\nu\ll 1$ the Hofstadter spectrum consists of well-separated (broadened) Landau levels. As a result the CF ground state is separated by an energy gap from  excitations.  This renders stability of the mean-field ansatz against small fluctuations. Notice that the CF spectrum is gapless at $\nu=1/2$, Fig.~\ref{fig-Hofstadter}, (indeed, flux per cell is $\Phi=2\pi$ and may be gauged away), suggesting that the mean-field ansatz may be inapplicable (at least not in the form adopted above). 

To conclude, we considered the nature of the ground state and low-energy excitations of repulsive bosons on lattices with 
moat bands. The  optical lattices with appropriate characteristics have been reported \cite{exp1,exp2,exp3,exp4} very recently, opening a way for experiments on cold bosonic atoms with moat dispersion. We have shown that at small filling factors the expected ground state is not BEC, but is rather a filled 
LLL of composite fermions. The excitations are gapped and have fermionic statistics. This manifests itself in discontinues 
jumps of the chemical potential at fractional filling fractions, Eq.~(\ref{eq-density-quantized}).


We are grateful to V. Galitski, D. Huse and O. Starykh  for useful discussions. This work was supported by DOE contract DE-FG02-08ER46482.






\begin{thebibliography}{1}
\bibitem{foot1} For three distinct $t_1,t_1',t_1''$ and  $t_2,t_2',t_2''$ the Hamiltonian has the structure of (\ref{H1}),
if $t_1t_2=t_1't_2'=t_1''t_2''$.
\end{thebibliography}
\end{document}